\documentclass[twocolumn, deluxetables]{aastex62}
\usepackage{amsmath}
\usepackage[caption=false]{subfig}
\usepackage{multirow}
\graphicspath{{./}{figures/}}

\shorttitle{GRS 1915+105}
\shortauthors{Balakrishnan et al.}

\newcommand{\grs}{GRS 1915$+$105}

\begin{document}

\title{The Novel Obscured State of Stellar-mass Black Hole GRS 1915+105}

\correspondingauthor{M.~Balakrishnan}
\email{bmayura@umich.edu}

\author{M.~Balakrishnan}
\author{J. M. Miller}
\author{M.T.~Reynolds}
\author{E.~Kammoun}
\author{A.~Zoghbi}
\author{B.E.~Tetarenko}
\affil{Department of Astronomy, The University of Michigan, 1085 S. University Ave., Ann Arbor, MI, 48109, USA}




\begin{abstract}
GRS 1915$+$105 is a stellar-mass black hole that is well known for exhibiting at least 12 distinct classes of X-ray variability and correlated multi-wavelength behavior.  Despite such extraordinary variability, GRS 1915$+$105~ remained one of the brightest sources in the X-ray sky.  However, in early 2019, the source became much fainter, apparently entering a new accretion state.   Here, we report the results of an extensive, year-long monitoring campaign of GRS 1915$+$105 with the \textit{Neil Gehrels Swift Observatory}.  During this interval, the flux of GRS 1915$+$105 gradually diminished; the observed count rate eventually dropped by two orders of magnitude.  Simple but robust spectral fits to these monitoring observations show that this new state results from the combination of a dramatic and persistent increase in internal obscuration, and a reduced mass accretion rate.  The internal obscuration is the dominant effect, with a median value of $N_{H} = 7\times 10^{23}~{\rm cm}^{-2}$.  In a number of observations, the source appears to be Compton-thick.  We suggest that this state should be identified as the "obscured state," and discuss the implications of this new (or rarely observed) accretion mode for black holes across the mass scale.
\end{abstract}



\section{Introduction} \label{sec:intro}
 Microquasars are stellar-mass black holes that replicate some of the behaviour of supermassive black holes on accessible time scales, allowing observers to better understand quasars and Active Galactic Nuclei (AGN) \citep{1999ARA&A..37..409M}. \grs \ is a well-studied and well-known microquasar due to its interesting properties: apparently superluminal radio jets \citep{1994Natur.371...46M}, distinctive X-ray variability classes \citep{2000A&A...355..271B}, high frequency QPOs \citep{2006csxs.book..157M}, and ``heartbeats'' \citep{2016ApJ...833..165Z}. The complex variability has been attributed to disk instabilities \citep[e.g.][]{2006MNRAS.372..728M} associated with the high accretion rate.  Indeed, since its discovery in 1992, \grs~ has persistently been one of the brightest sources in the X-ray sky.

In July of 2018, \grs \ unexpectedly started dimming: \citealt{2018ATel11828....1N} observed the lowest soft X-ray flux of \grs \ in the 22 years of monitoring with RXTE/ASM and MAXI/GSC, reporting a 2-10 keV flux of $2.4 \times 10^{-9} \rm \ erg \ cm^{-2} \ s^{-1}$ and an unabsorbed flux of $4.3 \times 10^{-9} \rm \ erg \ cm^{-2} \ s^{-1}$.  The broad similarities between this flux decline and the decay of other transient X-ray binary outbursts led \citealt{2018ATel11828....1N} to conclude that the long outburst of GRS~1915$+$105 was ending.  However, weekly observations with NICER then showed that GRS~1915$+$105 appeared to stabilise to a flux of 0.08 Crab. In April 2019, there was a sudden evolution in the source: the flux dropped to $1.2 \times 10^{-9} \rm \ erg \ cm^{-2} \ s^{-1}$ (0.5-10.0~keV), and QPOs that had previously been detected either halted or were no longer detectable \citep{2019ATel12742....1H}.  Contemporaneous Chandra observations found that the spectrum of GRS 1915$+$105 was marked by a number of strong, ionized absorption lines, potentially consistent with a slow, dense disk wind \citep{2019ATel12743....1M, 2020arXiv200707005M}.

In order to better understand this behavior, we launched a year-long X-ray monitoring campaign with the \textit{Neil Gehrels Swift Observatory} \citep{2004ApJ...611.1005G} (hereafter \textit{Swift}).  The cadence and sensitivity afforded by \textit{Swift} are ideal for tracing the outbursts of X-ray transients \citep[see, e.g.][]{2013HEAD...1312623R} ; particularly when an outburst decays to low flux levels, \textit{Swift} is the only means of obtaining monitoring that enables basic spectral modeling.  Through this program, we have previously reported evidence that GRS~1915$+$105 is sometimes Compton-thick in its obscured state \citep{2019ATel12771....1M, 2020arXiv200707005M}, and that the column density remains persistently high \citep{2019ATel12848....1B}.

During our \textit{Swift} monitoring campaign, a number of strong X-ray and radio flares were detected from GRS~1915$+$105 \citep[see, e.g.,][]{2019ATel12773....1M, 2019ATel12855....1T, 2019ATel13304....1T, 2020ATel13478....1T}.  These flares clearly signal jet activity, and they also clearly signal that the mass accretion rate onto the black hole remained relatively high despite the low flux.  This is consistent with flux from the central engine becoming heavily obscured.

In this paper, we present a comprehensive spectral analysis of 118 \textit{Swift} spectra obtained during the dim-variable period of \grs; spanning from April 2019 through April 2020.  The heavily obscured spectra that we obtained from GRS~1915$+$105 are very similar to spectra observed in Seyfert-2 AGN and Compton-thick AGN, so we fit each spectrum with AGN-like models including internal obscuration, scattering, and distant reflection.  We traced the evolution of the model parameters across the novel obscured state, and examined correlations between the key model parameters.  We also examined the light curve of each observation in order to identify flaring episodes.  Section 2 details the steps taken to reduce the data and the instrumental set-up. In Section 3, we describe our analysis of the \textit{Swift} spectra. Section 4 details the results obtained in our modelling, including the evolution of model parameters and potential correlations.  In Section 5, we discuss the implications of the novel obscured state observed in GRS 1915$+$105.

\section{Data Reduction}
We analyzed a total of 122 {\it Swift}/XRT observations of \grs, obtained between April 2019 and April 2020, and available in the public NASA/HEASARC archive. The identification number, start time, duration, and number of photon counts of each observation are given in Table 1.    For consistency and accuracy, our analysis was restricted to {\it Swift} observations that were obtained in WT (Windowed Timing) mode. Observations obtained in Photon Counting mode appear to occasionally suffer from photon pile-up; since it is difficult to recover the intrinsic continuum in heavily obscured spectra, pile-up mitigation is unlikely to be fully robust. 

We processed all of the observations the using tools available in HEASOFT version 6.26. Initial event cleaning was carried out using the task \texttt{XRTPIPELINE}.  We created source and background regions that measured around 60 arcseconds.  Spectra and light curves were extracted using \texttt{XSELECT}.  Ancillary response files (ARF) were creating using \texttt{XRTMKARF}, and finally, the spectra were binned to a signal-to-noise ratio of 3.0 using \texttt{FTGROUPPHA}.  In each spectral fit, we used the default redistribution matrix file (RMF) for WT mode available in the \textit{Swift} calibration data base (CALDB version 20180710).

\textit{Swift} observed \grs \ 122 times with WT mode during the monitoring period. After binning, 15 observations did not have enough counts for spectral analysis at a signal-to-noise ratio of $S/N \geq 3$, leaving 107 useful observations. Some the remaining observations required additional considerations.  For the small number of observations consisting of multiple good time intervals (GTIs), where possible, we analyzed each GTI separately. This was important because the internal obscuration sometimes decreased or increased significantly within the course of a single day. In the cases where one or more segments had too few counts to allow for meaningful spectroscopy, the time-averaged spectrum was analyzed. We also observed two full X-ray flares, and two segments of X-ray flares. OBSIDs 00012178006 and 00012178037 captured full flares.  OBSIDs 00034292162 and 00012178019 captured the rise of a flare during the exposure.  For these observations, the flare and steady emission were analyzed separately. During fitting, we noticed that four observations had the same number of degrees of freedom, and very low $C/\nu$ values, where C refers to the Cash statistic, explained in the following section. The fits also returned a null hypothesis value of 1, and the spectra have therefore been omitted. 

This leaves us with 96 untouched WT-mode observations, seven observations that were split into two segments, and four observations that were separated into flare/non-flare spectra. Therefore, we present results from a total of 118 spectra. 

\section{Analysis} 

All spectral fits were made using XSPEC version 12.10.1
\citep{1996ASPC..101...17A}.  In WT mode, fits are sometimes extended down to 0.6 keV \citep[e.g.,][]{2007ApJ...666.1129R, 2013HEAD...1312623R}.  In the case of GRS~1915$+$105, there are very few counts below 1 keV, so we adopted this energy as our lower fitting bound. We adopted an upper fitting bound of
10 keV, the upper limit of the calibrated XRT band.  The fits
minimized the Cash statistic, or ``C'' statistic \citep{1979ApJ...228..939C}. It must be noted that in reality \texttt{XSPEC} uses a slightly modified version of the Cash statistic when background spectra with Poisson noise is added called the W statistic. Please see \citealt{1996ASPC..101...17A} for further details. The errors reported in this work are 1 $\sigma$ errors.  The errors on derived quantities were calculated using standard error propagation, and are also 1 $\sigma$ confidence errors.

Despite a relatively modest number of counts, it is clear that the
spectra differ considerably from those typically obtained in low-mass
X-ray binaries.  These differences are illustrated in Figure 1,
using the spectrum from OBSID 00034292201.  Stellar-mass black holes
in the ``low/hard'' state exhibit power-law continua (see \citealt{2013ApJ...775L..45M} for modeling of GRS 1915$+$105 in a typical ``low/hard'' state and \citealt{2016ApJS..222...15T} for behaviour of stellar mass black holes in this state).
Fits to OBSID 00034292201 with a plausible Galactic absorbing column
($N_{H} = 4.35\times 10^{22}~{\rm cm}^{-2}$ and reasonable power-law
index ($\Gamma = 2$) completely fail to fit the spectrum.
Additional obscuration at low energy must be the result of an
additional internal column of cold or low-ionization gas.  The
continuum in the Fe~K band is altered by very strong reflection, as
indicated by an Fe~K emission line at 6.4~keV and edge at 7.1~keV.

As noted by \citealt{2020arXiv200707005M}, this kind of spectrum is very
similar to those obtained from heavily obscured black holes in
Seyfert-2 or even Compton-thick AGN.  We therefore adopted a
simplified version of a set of models recently employed by \citealt{2020ApJ...901..161K} to study a sample of Seyfert-2 AGN. The model consists of an
additional internal column density component acting on the direct continuum, a scattered continuum component that can have a different column
density, and unobscured cold reflection (via ``pexmon,'' see Nandra et
al.\ 2007).  In \texttt{XSPEC} parlance, this model can be written as follows:
\texttt{phabs$_{los}$ $\times$ (phabs$_{1}$ $\times$ cutoffpl +
const $\times$ phabs$_{2}$ $\times$ cutoffpl $+$ pexmon)}.  

The \texttt{phabs$_{los}$} component describes line-of-sight absorption
in the Milky Way, and its column is fixed at $N_{H} = 4.35\times
10^{22}~{\rm cm}^{-2}$ \citep{2009ApJ...706...60B}.  The \texttt{phabs$_{1}$} component describes
the internal obscuration and it acts directly on the cut-off power-law
continuum.  In all fits made in this work, the cut-off energy was
fixed at $E_{cut} = 30$~keV \citep[e.g.,][]{2013ApJ...775L..45M}.  In initial tests, we found that the power-law index was poorly constrained but always consistent with Gamma = 2.0; this value was therefore frozen in all of our fits. The \texttt{phabs$_{2}$} component group allows for scattered emission that may suffer different obscuration; the \texttt{const} factor is
allowed to vary between zero and unity.  The \texttt{pexmon} component
is fit as a ``reflection only'' component, with its continuum parameters
(the photon power-law index, cut-off energy, and flux normalization)
linked to the values in the \texttt{cutoffpl} component). The reflection fraction was bounded to $-5 \leq R \leq 0$. The negative sign only indicates that ``pexmon'' was fit as a pure
reflection spectrum, without its own continuum.  A value of $R = -1.0$
corresponds to the reflector covering the full $2\pi$ steradians of
the sky, as seen from the emitter. The
reflection inclination was fixed to $\theta = 72^{\circ}$ (as per
\citealt{2013ApJ...775L..45M}; note, however, that the spectrum is insensitive
to this parameter), and the abundances were frozen at solar values.
The reflection fraction was allowed to vary. 

We decided this model was superior to others after in-depth investigation. Model superiority was determined using the Akaike Information Criterion (AIC) in a form seen in \citealt{2016MNRAS.461.1642E}, where it is corrected for the bias arising from the finite size of the sample.

\begin{equation*}
    \rm AIC_C = 2k - 2 C_L + C + \frac{2 k (k+1)}{N- k - 1},
\end{equation*}

\noindent where $\rm C_L$ is the constant likelihood of the true hypothetical
model, and does not depend on either the data or tested models, C is
the Cash statistic value, k is the number of free model parameters,
and N is the number of data points. When taking the difference of the AIC of two models, the constant likelihood factor cancels out. If $\rm \Delta AIC_C < 2$, then the models fit the data equally well. Below, we describe alternative models that were investigated, along with approximate values of the associated $\rm \Delta AIC_C$.\\

\noindent$\bullet$ Ionized obscuration ($\rm \Delta AIC_C > 15$): In one trial, we changed the $\texttt{phabs}_1$
component to a \texttt{zxipcf} component, which uses a grid of \texttt{XSTAR} photionised absorption models for the absorption and assumes the gas only covers some fraction of the source. This allowed us to examine
the possibility that the internal obscuration is ionized, and that it
may only partially cover the central engine.  In our tests,
\texttt{zxipcf} did not improve the fit; in many cases, poorer fits
resulted, with $\rm \Delta AIC_C > 15$. \\

\noindent$\bullet$ Diffuse, ionized emission ($\rm \Delta AIC_C > 15$): We also tried adding an \texttt{apec} component \citep{2012ApJ...756..128F}.
This would capture emission from diffuse, ionized gas, likely
dominated by Fe XXV and Fe XXVI lines \citep[see, e.g., ][]{2020arXiv200707005M}.  The addition of this component also tended to {\it increase}
the C-statistic.  At least at the resolution and sensitivity afforded
by the XRT monitoring spectra, diffuse emission is not strongly
required.\\

\noindent$\bullet$ Constraining the scattering component ($\rm \Delta AIC_C > 15$): It is reasonable to expect that the contribution of the scattered emission may be fairly minor. However, fits to the spectra with the value of the constant constrained to be less than 0.1 were significnantly worse. Despite their modest sensitivity, the data prefer a degree of scattered emission, and values of the constant factor that are a few times higher than this limit. \\

\noindent$\bullet$ More limited reflection fraction ($2 < \rm \Delta AIC_C < 8$): At the beginning of our analysis, it seemed appropriate to assume that the obscuring gas formed a disk around the central engine, given that disk winds in stellar-mass black holes are generally equatorial \citep{2006Natur.441..953M, 2012MNRAS.422L..11P}.  We therefore bounded the reflection fraction to stay between $-1 \leq R \leq 0$. This resulted in 40\% of the observations hitting the limit of $R = -1$. Given the extremity of this novel state of GRS 1915$+$105, a new lower limit of $-5 \leq R \leq 0$ was adopted.  Improved fits were then achieved, potentially indicating that the disk wind reached large scale heights throughout the course of our observation campaign. \\ 

Figure 1 shows the spectrum obtained in OBSID 00034292201, fit
with our model.  An excellent fit is obtained ($C/\nu = 64.6/64$).
The additional absorbing column at low energy, the Fe~K emission line,
and the Fe~K edge are all modeled well.  Moreover, the internal obscuration and reflection parameters can be constrained, and
correlated.

Figure 2 helps to illustrate the features and action of this model.
We searched for archival XRT spectra obtained in the ``low/hard'' state,
with an unabsorbed flux comparable to spectra in the obscured state.
OBSID 00030333242, obtained on March 23, 2016, is shown in Figure 2.
Its spectrum can be fit with a simple power-law continuum, modified by
absorption in the ISM.  The spectrum of OBSID 00034292186, obtained on
June 24, 2019, can be fit with the same power-law index {\it and}
normalization.  However, the more recent spectrum requires a very
large internal column density and distant reflection, as per the
models outlined above.  Clearly, the internal obscuration is not only
strong, but the defining characteristic of this source state, making
it easy to measure even in our snapshot spectra.

In addition to measuring key parameters, we
also obtain absorbed and unabsorbed flux values of the central engine.  Since modeling
highly obscured spectra can be particularly difficult, we only report
fluxes from the 1--10~keV band used for spectral fitting.  We report
the unabsorbed flux of the central engine ($F_{cent, unabs}$) as the flux from the main
\texttt{cutoffpl} component only, after removing both internal and
line-of-sight obscuration.  The flux from the scattered and reflected
components is tracked separately.

\begin{figure*}[ht]
\centering
\includegraphics[width=0.48\textwidth, angle=270]{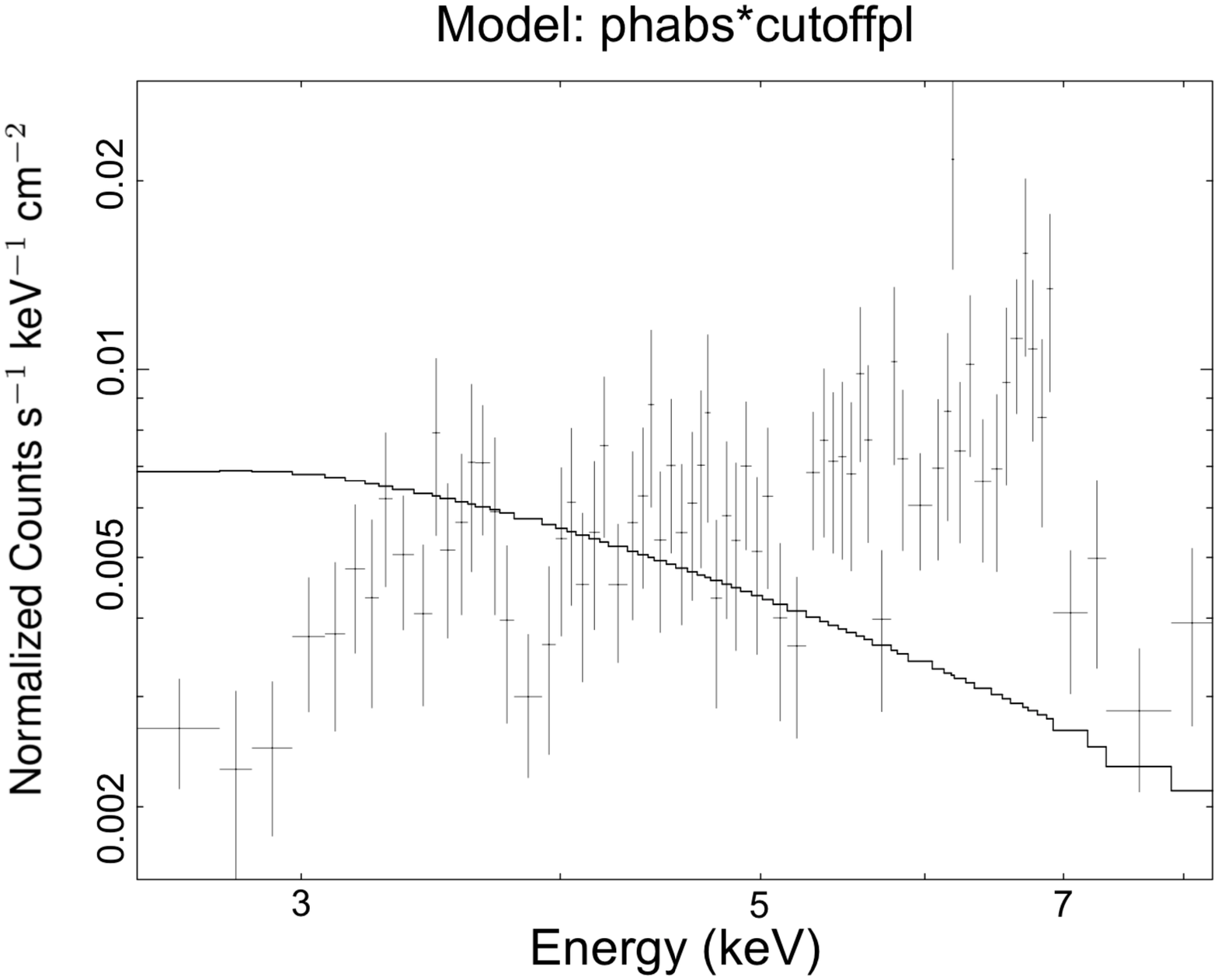}
\includegraphics[width=0.48\textwidth, angle=270]{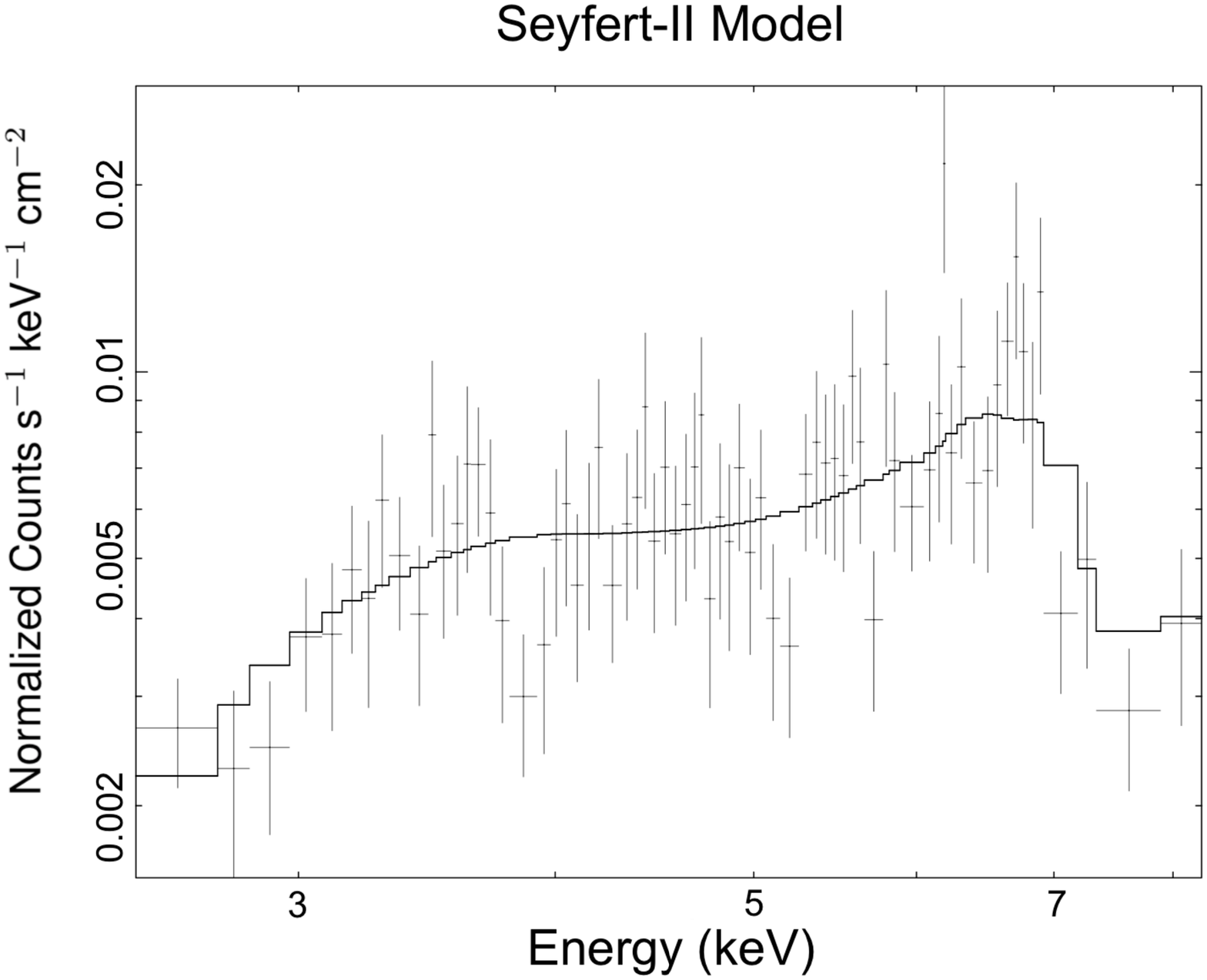}
\vspace {1mm}
\caption{A \textit{Swift}/XRT spectrum of GRS 1915$+$105 in the obscured state (OBS 00034292201).  LEFT: The spectrum is fit with a simple model: \texttt{phabs} $\times$ \texttt{cutoffpl}.  The Galactic line-of-sight absorption is frozen to $N_H = 4.35 \times 10^{22} \rm cm^{-2}$ and the cutoff energy was frozen to $E_{\rm cut} = 30$ keV.  The photon index was allowed to vary within a plausible range ($1.4 < \Gamma < 2.2$).  The model fails at low energy owing to increased obscuration, and in the Fe~K band owing to reflection (indicated by an Fe~K emission line at 6.4~keV and Fe~K edge at 7.1~keV).  This model results in an unacceptable fit (C / $\nu$ = 323.17/68).  RIGHT:  The spectrum fit with a model including internal obscuration, scattering, and reflection: \texttt{phabs$_{low}$ $\times$ (phabs$_{1}$ $\times$ cutoffpl + 
const $\times$ phabs$_{2}$ $\times$ cutoffpl $+$ pexmon)}. This model achieves an excellent fit (C / $\nu$ = 64.63/64).  }
\end{figure*}

\begin{figure*}[htb!] 
\centering
\includegraphics[width=0.49\textwidth, angle=270]{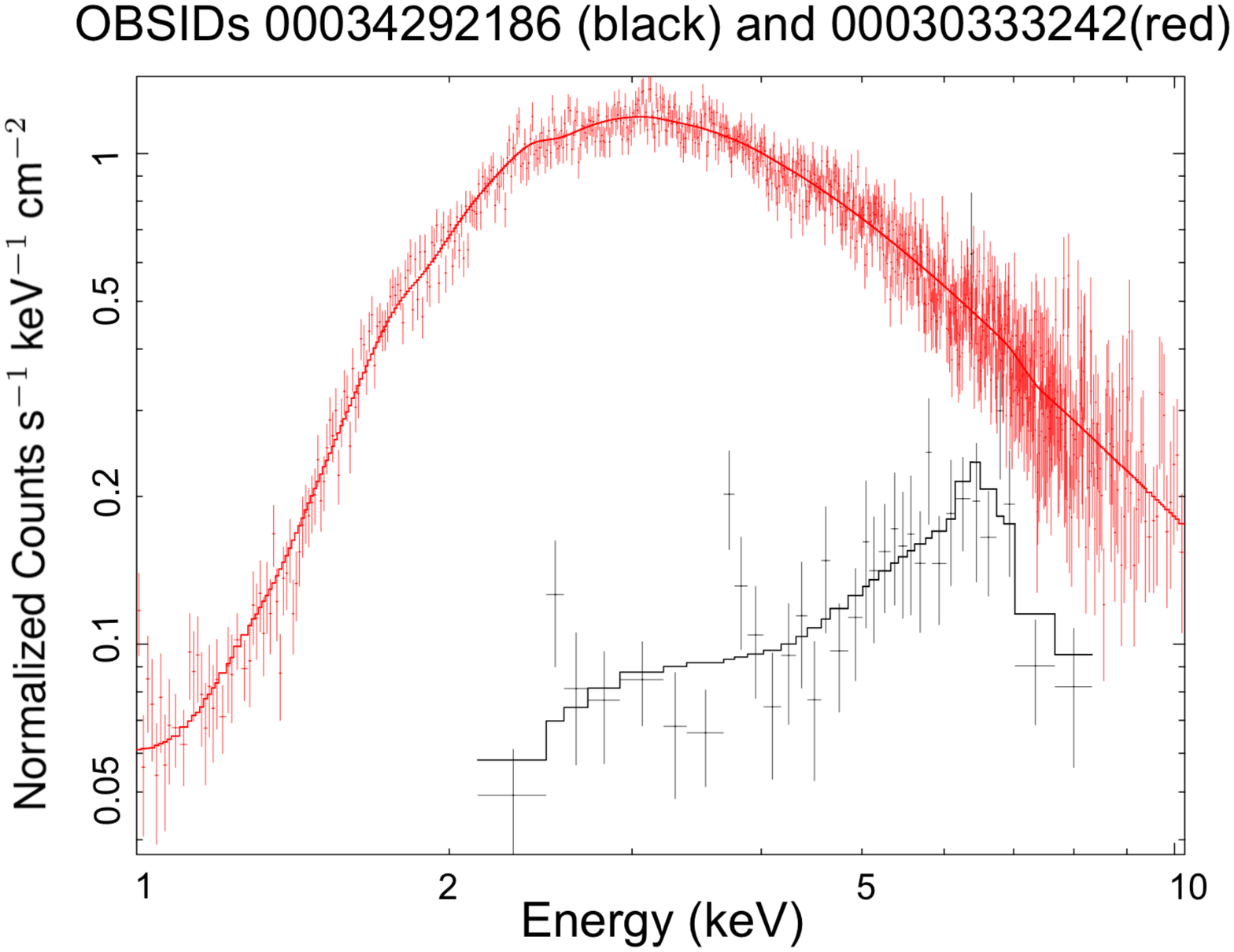}
\includegraphics[width=0.49\textwidth, angle=270]{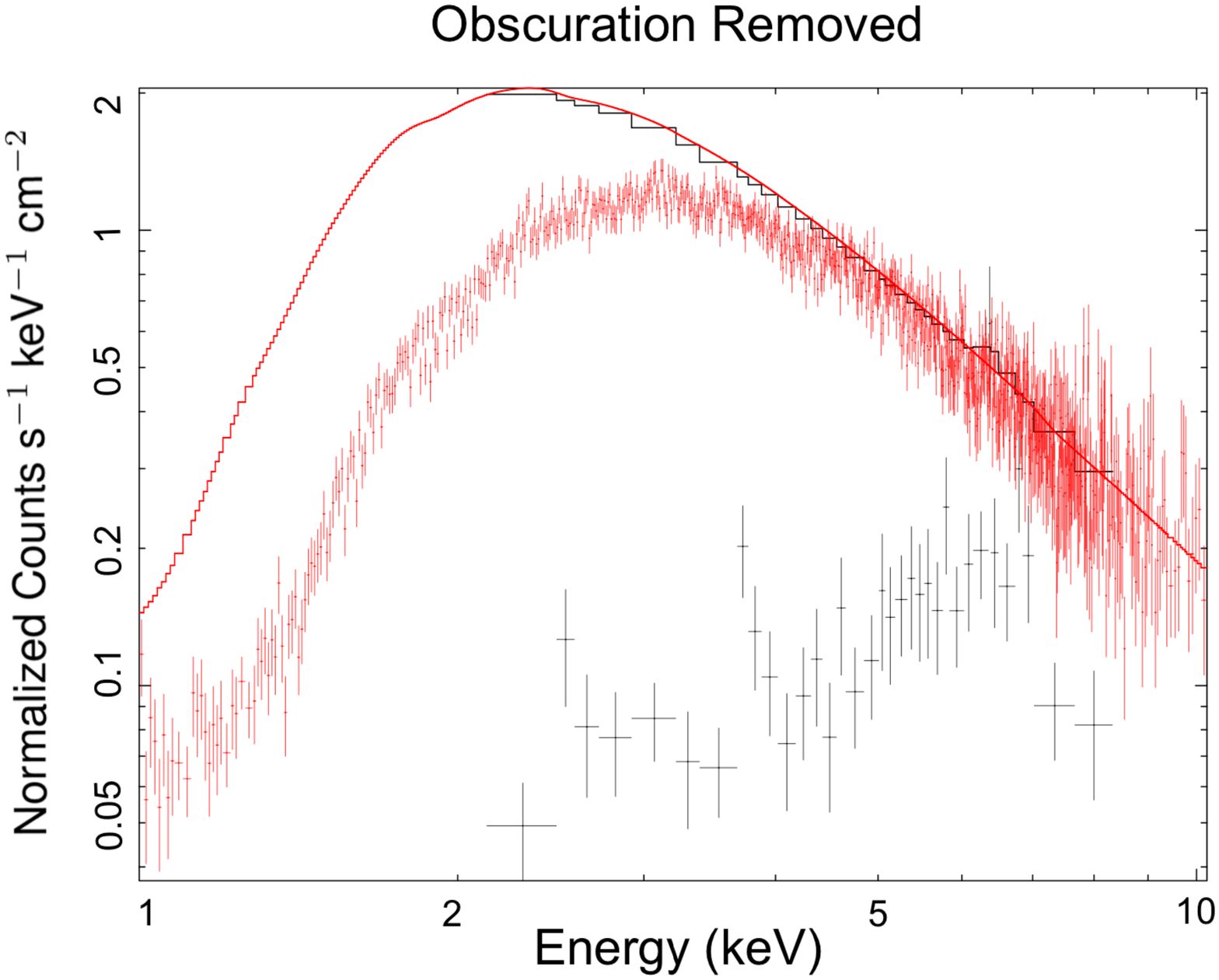}
\vspace {1mm}
\caption{Two spectra of GRS 1915$+$105 with comparable Eddington fractions, taken twenty-six months apart.  OBS 00034292186 (obtained during the obscured state, on 24 June 2019) is shown in black and 00030333242 (obtained during a normal ``low/hard'' state, on 23 March 2016) is shown in red.  The obscured spectrum has an {\rm unabsorbed} flux of $1.4 \times 10^{-7}$ erg/cm$^2$/s while the normal spectrum has a an unabsorbed flux of $2.6 \times 10^{-7}$ erg/cm$^2$/s. The normal spectrum was binned to a signal-to-noise ratio of 3.0, like the other observations.  The similar flux levels and spectral shapes yield very different observed spectra owing to very high internal obscuration in GRS 1915$+$105 throughout 2019.}
\end{figure*}

\section{Results}

Table 2 lists the continuum parameters measured from fits to the 118 \textit{Swift}/XRT spectra of GRS 1915$+$105 in its obscured state.
Figure 3 depicts the evolution of the key model parameters, as
well as the observed and unabsorbed flux of the central engine, the ratio of those fluxes,
and the Eddington fraction of the source.  The Eddington fraction was calculated using the 1-10 keV unabsorbed flux of the central engine ($F_{cent, unabs}$), where we assumed a distance of 8.2 $\substack{+2 \\ -1.6}$ kpc \citep{2014ApJ...796....2R}.  The flares are shown in each plot with orange markers.  

The top panel in Figure 3 shows the time evolution of the most important quantity measured in our fits, the internal column density that acts on the direct continuum ($\texttt{phabs}_{1}$).  In this panel, it is clear that $\texttt{phabs}_{1}$ is initially low, only twice as high as the value in the ISM.  Over a period of 40 days, the column density increases until it is consistent with being Compton-thick on MJD 58619 (a red horizontal line marks the column density at which the obscuration becomes Compton-thick, $N_{H,1} = 1.5\times 10^{24}~{\rm cm}^{-2}$).  Thereafter, the internal column density
is not less than $N_{H,1} \simeq 2.3 \times 10^{23}~{\rm cm}^{-2}$, and
it is typically much higher over the remaining span of approximately
300 days.  In total, 10\% of the observed spectra are consistent with
being Compton-thick.  These episodes are distributed fairly
randomly across the monitoring period.

The column density associated with scattered emission, $N_{H,2}$, also
varies.  However, it is always lower than the primary obscuring column
$N_{H,1}$.  The reflection fraction is weakly constrained by the data.  Although most values stay between $-1 < R < 0$, there are several instances of high reflection fractions, potentially signaling that the obscuring geometry has a significant vertical extent.  In other observations, the reflection fraction is formally consistent with zero.  This nominally indicates that the reflector is potentially highly variable.  

The observed and unabsorbed flux of the central engine, their ratio, and the inferred
Eddington fraction are also shown in Figure 3.
The measured values are also listed in Table 3, which also lists the flux from each component of the spectral model. These parameters vary by orders
of magnitude across the monitoring period, with no clear pattern.  The
most important and robust result is simply that obscuration reduces
the observed flux by 1--2 orders of magnitude.  It is the most
important factor in making GRS 1915$+$105 faint through 2019 and 2020;
a reduced mass accretion rate is only a secondary factor (also see
Figure 2).  Indeed, in some observations -- {\it not} the flares,
remarkably -- GRS 1915$+$105 is nominally super-Eddington.

\subsection{Correlations}
In Figure 4, we examine correlations between $N_{H,1}$, $N_{H,2}$, the reflection fraction, and the unabsorbed central engine flux.  Logarithmic quantities were considered as the values of some parameters spanned orders of magnitude.  Observations for which $N_{H,2} \leq 10^{22}~{\rm cm}^{-2}$ were omitted as this column density is the lower limit of what can be detected in addition to the interstellar column density.  The XRT sensitivity limit is $2 \times 10^{-14}$ erg cm$^{-2}$ s$^{-1}$ \citep{2004SPIE.5165..201B}, and therefore observations with reflected flux levels below this limit were omitted. In order to concentrate on the fundamental properties of the obscuration and accretion flow, we also omitted the spectra containing flares.

Spearman Rank-Order correlation coefficients ($r_s$) between the parameters and corresponding p-values, the probability there is no underlying correlation, were calculated using Monte-Carlo sampling over the uncertainty of the data to account for errors. Correlation coefficients close to +1 or -1 indicate high correlation, while a coefficient of 0 indicates there is no correlation. With a significance level of $p < 0.05$, three main trends arise, presented in Figure 4. Correlations between parameters that were not significant have been omitted in the interest of space and clarity.  We do note that the individual flux components are  positively correlated; this is expected since the scattered and reflected components are still driven by the central engine, and these correlations are also omitted because they are easily anticipated and explained. It is clear that the reflection fraction, internal obscuration, scattered gas, and unabsorbed central engine flux are related. The internal obscuration is positively correlated with the unabsorbed central engine flux ($r_s$ = 0.25) and the scattered gas ($r_s$ = 0.58), while being negatively correlated with the reflection fraction ($r_s$ = -0.30).  This suggests that the gas that is out of the line of sight is connected to the gas in our line of sight (possibly pointing to an axially symmetric disk wind) and that stronger flux from the central engine causes an increase in gas that builds up equatorially. The correlation between internal obscuration and unabsorbed central engine flux is weak but significant, which may suggest that the flux is sometimes coupled with the absorber, but not permanently.

\subsection{Flares}

As noted previously, four observations captured flares.  In two of these observations, the rise {\textit{and}} fall of the flare was captured.  Closer inspection of OBSIDs 00012178006 (MJD 58807) and 00012178037 (MJD 58935) revealed a more complex flaring structure; these observations are shown in Figure 5.   The high signal to noise ratio in these light curves allowed us to calculate a decay time for the three flares seen in OBSID 00012178006 and the four flares seen in OBSID 00012178037.  Using \texttt{scipy.odr} \citep{1990...ODR} in python, we fit the decay of each flare with an exponential function to get a characteristic time scale.  Five of the flares have decays consistent with $\tau = 20$~s, but shorter flares are also observed, leading to the conclusion that there is not a single characteristic time scale that might help to identify a flaring mechanism. 

Nevertheless, it is useful to compare the flaring time scales to plausible dynamical ($t_{dyn} = \sqrt{\frac{r^3}{G M}}$), thermal ($t_{thermal} = \frac{t_{dyn}}{\alpha}$), and viscous ($t_{visc} = t_{thermal} \Big( \frac{r}{h_d} \Big)^2 $) time scales in the disk.  For an assumed viscosity parameter $\alpha = 0.1$, characteristic scale height $\frac{h_d}{r} = 0.1$, and mass of $12.4 M_\odot$ \citep{2014ApJ...796....2R}, the viscous time scale at a radius of $r = 100~GM/c^{2}$ is $t_{visc} = 27$~s, comparable to the longest flares observed in our monitoring program.  The dynamical and thermal time scales are much shorter, of course.  This may nominally suggest that viscous fluctuations at intermediate radii play a role in the observed flaring.  However, it is not clear how most of the X-ray flux could be generated at $r = 100 GM/c^{2}$, rather than deeper within the potential.  Especially given that radio flaring was observed contemporaneously with X-ray flaring  \citep{2019ATel12773....1M, 2019ATel12855....1T, 2019ATel13304....1T, 2020ATel13478....1T}, it is more likely that flares are generated in a compact corona, which may itself be the base of the jet.  This does not preclude disk time scales partially determining the flaring time scales.

In Figure 6, column densities, reflection fraction, fluxes, flux ratio, and Eddington fraction are plotted for the flares.  Each flare was segmented into three parts, from low to high flux. The orange, purple, and red points refer to high, medium, and low flux sections of the flares, respectively. We find that the spectral parameters of the flares do not show any clear trends. For example, a higher central engine flux does not necessarily result in lower obscuration values, or vice versa. These parameters are also listed in Table 4.  Some flares may help to clear the internal obscuration (e.g., OBSID 0034292162), while others do not (e.g., OBSID 00012178006).  Again, it is possible that the observed flares are not the result of a single physical process, but rather a manifestation of different processes. 

In Figure 7, histograms of the primary obscuration column density, scattered column density, and reflection fraction are shown.  The first eight observations are considered separately from subsequent observations that are more typical of the obscured state.  Similarly, flaring segments (see below) are also considered separately.  The initial observations show lower reflection fractions, potentially consistent with standard reflection from the accretion disk.  Later observations extend to much larger reflection fractions, indicative of a geometry with greater vertical extent.

\section{Discussion}

The extraordinary variability of the stellar-mass black hole GRS~1915$+$105 has been studied extensively, but in early 2019 it entered a new accretion state that we call the “obscured state”, characterized by a low observed flux driven by increased internal obscuration. The  spectra bear strong similarities to those typically observed in highly obscured Seyfert-2 AGN. Using a spectral model based on those sources, we find that the internal obscuration increased  at the onset of the monitoring period and then remained very high, reducing the observed flux by an order of magnitude.  Analysis of relevant spectral parameters revealed three main trends: (1) a positive correlation between internal obscuration and unabsorbed flux of the central engine, (2) a positive correlation between the internal obscuration and scattered gas, and (3) a negative correlation between reflection fraction and the internal obscuration. In the following section, we discuss the nature of the obscuring gas, before ending with comparisons to other astronomical objects and proposed future studies.

The obscured state appears to be driven by obscuration of the central engine, not by a drop in the mass accretion rate onto the black hole.  The first few observations showed a relatively low internal obscuration column density (~$10^{23}$ cm$^{-2}$), but this value increased and stayed high throughout our monitoring campaign, with a median value of $7 \times 10^{23}$ cm$^{-2}$ (see Figure 3).  
The Eddington fraction, calculated using the unabsorbed direct flux from the central engine, remains relatively high during the monitoring period, with a median value of $L_{\rm Eddington}$ = 0.1. 
The Eddington fraction is not systematically higher early in the monitoring period, before the obscuration increases (see Figure 3).

Variations in, and correlations between, the column density and spectral components can help to reveal the nature of the obscuration and the geometry of the accretion flow.  Figure 4 shows that the obscuring column density increases when the central engine is more luminous (when the unabsorbed flux from the central engine is higher). This signals that the column density and the mass accretion rate are linked, consistent with some expectations of accretion disk winds. \citealt{2020arXiv200707005M} recently analysed three Chandra spectra of GRS~1915$+$105 in the obscured state and found evidence of a failed disk wind, and suggest that this eventually envelops the central engine.  The relationship observed between the obscuring column density and central engine flux is consistent with an accretion flow that is struggling to eject a disk wind, and largely failing.  

\textbf{\citealt{2020arXiv200707005M} also notes the presence of ionization lines as the source transitioned to the obscured state, and neutral and ionized emission lines once the source was clearly within the obscured state.  Most of our \textit{Swift} monitoring was obtained later, during the main part of the obscured state.  Neither the \textit{Chandra} data nor the \textit{Swift} data show evidence of ionized absorption in the obscured state.  This supports other evidence that the accretion flow evolved significantly, potentially including changes in covering factors and column density. The ionized emission lines detected in \textit{Chandra} and \textit{NICER} observations in the obscured state \citep{2020arXiv200707005M, 2020ApJ...902..152N} are not clearly detected in our \textit{Swift} spectra, though they cannot be ruled out.  \citealt{2020arXiv200707005M} attributes these lines to a failed wind, rather than disk reflection.}

Variability in the obscuring column density, reflected flux, scattered flux, and correlations between these parameters may reveal aspects of the wind geometry and its evolution.
Most observations had reflection fractions between -1 and 0, consistent with the expectations of an accretion disk and/or equatorial wind, but there were also several with higher reflection fractions. How can the higher reflection fractions be explained?  In principle, a warped disk could create higher reflection fractions, but the data likely point to a simpler explanation: the obscuring column and scattered flux are positively correlated (see Figure 4), indicating that the geometry is axially symmetric. Variations in the scale height of a cylindrical structure could lead to reflection fractions greater than unity. 

Figure 4 shows a negative correlation between the reflection fraction and the obscuring column density.  This means that the reflection fraction is highest at points when the measured obscuring column is not extremal, e.g., at points when the column is not Compton-thick (we find that approximately 20\% of spectra imply Compton-thick  obscuration.)
This may indicate that the wind settles into a more equatorial geometry when the gas is most dense, perhaps because forces that would serve to lift the gas above midplane and expel the wind are particularly overwhelmed (e.g., emergent magnetic fields might be overwhelmed by gas pressure).

In other states, the disk in GRS~1915$+$105 may be affected by a thermal instability \citep[e.g.,][]{2012ApJ...750...71N, 2016ApJ...833..165Z}.  Simulations predict that this would cause the scale height of the accretion flow to increase and decrease, as the disk empties and refills \citep{2007ApJ...666..368L}. Although we do not observe periodic variations, the behaviour of the thermal instability could potentially contribute to or drive the variations that we have observed.  Importantly, the behaviour described in \citealt{2007ApJ...666..368L} produces significant outflows dependent on the disk viscosity.

\textit{Swift} is a small telescope, and it makes short exposures.  We have not observed GRS~1915$+$105 in a bright phase, when the count rate might compensate for a small effective area and short exposure times.  Moreover, it can be difficult to uniquely decompose highly obscured spectra into different flux components, even when a fairly high level of sensitivity is achieved.  It is therefore worth briefly examining if the above inferences are likely to be robust.
Figure 8 plots confidence contours for $N_{H}$ and reflection fraction, for observations that were taken just one day apart (ObsIDs 00034292227 and 00034292228).  The contour plots were made using \texttt{STEPPAR} in \texttt{XSPEC}.  We stepped through column density values from $0$ to $400 \times 10^{22} \rm cm^{-2}$ in 200 steps and reflection fraction values from $-5$ to $0$ in 100 steps.  It is clear that the values for column density exclude each other, even up to their respective 2 $\sigma$ errors.

The rise and fall of two flares were captured in our light curves (see Figure 5). The flares may have different emission mechanisms and absorption properties, so they were analyzed separately from the spectra that contribute to the correlations in Figure 4. The evolution of the spectral parameters of the flares is shown in Figures 6 and 7.  The most important difference between the regular flux intervals and the flares is that the reflection fraction is not anti-correlated with the internal obscuration in the flaring intervals.  Instead, during the flares, the column densities and reflection fractions were simultaneously low. We do not find a relationship between the internal obscuration and source luminosity; However, in a flare better observed in \citealt{2020ApJ...902..152N}, which analyzed one flare of \grs \ in the obscured state, a strong positive correlation between partial covering fraction and source luminosity is observed. The X-ray flares are clearly linked to strong radio flares \citep[e.g.,][]{2019ATel12773....1M}, likely indicating a link to jet production.  If the X-ray corona is the base of the jet, and if its scale height is greater during flaring intervals, its flux may be less subject to obscuration and the reflecting gas would subtend a smaller fraction of the solid angle.  We did not detect enough flares, and did not achieve enough sensitivity within each flare, to strongly confirm or reject this hypothesis.

How does the obscured state compare to better-known states observed in stellar-mass black holes?   The median value of the Eddington fraction, $L/L_{Edd} \sim 0.1$, is relatively high and either consistent with the soft (or, thermal-dominant) state or the mixed states \citep{2006ARA&A..44...49R, 1991ApJ...374..741M, 1993ApJ...403L..39M}.  The heavy obscuration makes it particularly difficult to detect a weak and/or cool disk component in our spectra; however, the spectra are fully consistent with a simple power-law.  Very strong radio and X-ray flaring is inconsistent with the soft state, and is rare even in mixed states. In these respects, the continuum spectrum and variability of the obscured state of GRS~1915$+$105 are not fully consistent with definitions of standard X-ray states. 

Similar obscured states may exist in other stellar-mass black holes with low-mass companion stars.  However, they may previously have been missed owing to a combination of factors.  First, such states may be short in systems with short orbital periods.  GRS~1915$+$105 happens to be the stellar-mass black hole with the longest known orbital period (for recent popoulation studies, see \citealt{2016ApJS..222...15T, 2016A&A...587A..61C}).  Second, the \textit{Swift}/XRT is more sensitive than prior instruments that monitored X-ray binary outbursts, and its pass band extends to lower energies. It is possible that prior monitoring was largely insensitive to changing internal obscuration.  Third, and most importantly, if obscured states are driven by failed winds, obscured states may be partly a matter of viewing angle.  Disk winds are equatorial \citep[see, e.g., ][]{2006Natur.441..953M, 2012MNRAS.422L..11P}, so obscuration would be most prominent in systems viewed at a high inclination.  In systems viewed closer to the pole, strong, narrow, Fe K lines should be seen, but these may be mixed with relativistic reflection from the inner disk.

GRS~1915$+$105 is not the only stellar-mass black hole where in heavy internal obscuration has been observed.  However, the detection of heavy obscuration in our observations of GRS~1915$+$105 is particularly important because it carries consequences for black holes across the mass scale.  The obscuration observed in GRS~1915$+$105 must come from
the accretion flow, not the low-mass companion star.  This situation differs markedly from heavy obscuration in systems like Cygnus X-3 and V4641~Sgr  \citep[see, e.g., ][]{2014ApJ...786L..20M}.  These sources have Wolf-Rayet and B stellar companions, respectively, and the obscuration is likely the result of the massive companion wind.  It is not clear that wind-fed binaries are equally good analogies for accretion onto supermassive black holes. Therefore, future studies of this obscured state should target neutron stars or black holes that have a low mass companion, a high inclination, and a long orbital period. 

V404 Cyg is another low-mass X-ray binary that harbors a black hole in a wide orbit.  It has also displayed evidence of strong and potentially Compton-thick internal obscuration, but only at points when the source was significantly super-Eddington \citep[see, e.g., ][]{2015ApJ...813L..37K, 2017MNRAS.468..981M}.  Few X-ray binaries have shown super-Eddington phases, and the same is true of AGN \citep{2009ApJ...696..891H}.  It is unclear if the obscured state in GRS 1915$+$105 and the brief Compton-thick episodes observed in V404 Cyg are fundamentally similar but and differ only in degree.  However, the fact that obscured states are possible at very different mass accretion rates in stellar-mass black holes signals that obscuration in Seyfert-2 AGN and Compton-thick AGN may not only be a matter of viewing angle, but also a matter of evolutionary phase and the nature of the innermost accretion flow.  

We thank Director Brad Cenko and the \textit{Swift} mission planning team for executing this monitoring program, especially at a time when in-person operations and meetings became impossible. We also thank the anonymous referee for comments that improved the manuscript.
JMM acknowledges helpful conversations with Tahir Yaqoob and Richard Mushotzky. Mayura Balakrishnan thanks Lia Corrales for helpful conversations.

\begin{figure*}[ht]
\includegraphics[height=0.9\textheight]{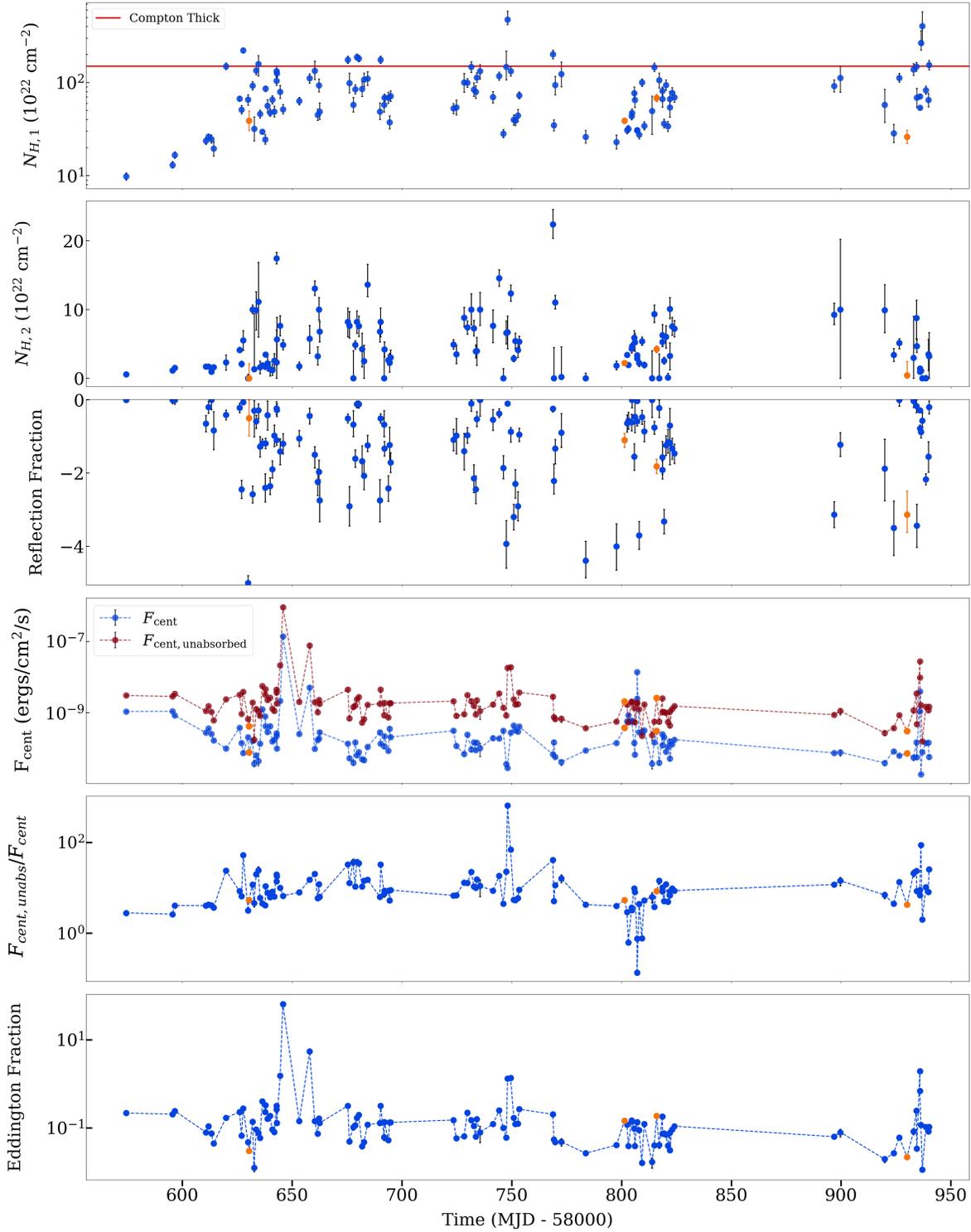}
\vspace {1mm}
\caption{Continuum parameters and fluxes are plotted for all spectra. In descending order, we plot values for the two \texttt{phabs} column densities, the reflection fraction, the 1-10 keV absorbed and unabsorbed fluxes of the central engine ($F_{cent}$ and $F_{cent, unabs}$), the unabsorbed to absorbed flux ratio ($F_{cent, unabs}/F_{cent}$), and the Eddington fraction calculated using the 1-10 keV unabsorbed flux. The points in orange correspond to spectra where a flare was observed in the light curve. The red line in the first panel denotes the measure of one Thompson optical depth. }
\end{figure*}

\begin{figure*}[ht]
\centering
\includegraphics[width=0.7\textwidth, angle=270]{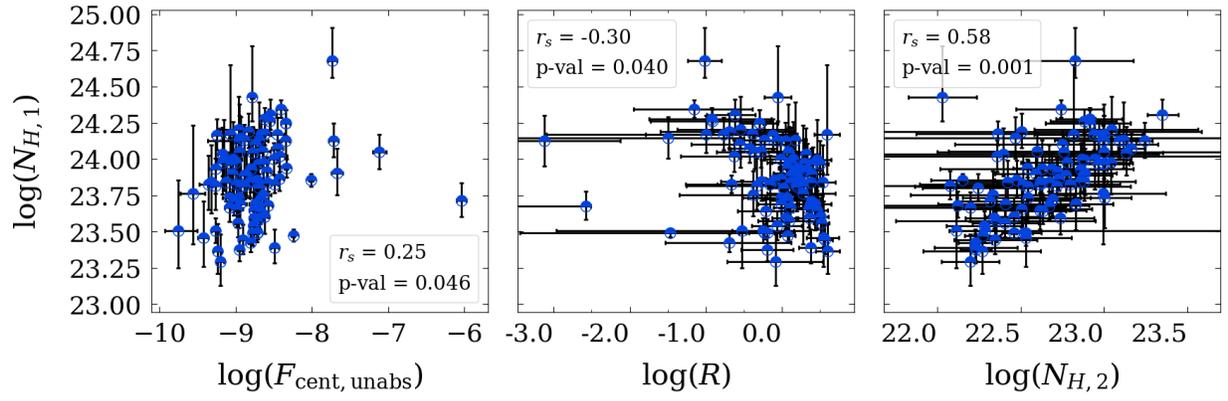}
\vspace{1mm}
\caption{Correlations between internal obscuration and $F_{\rm unabsorbed, central}$, reflection fraction, and scattered gas are shown. All values are in logspace. The Spearman Rank-Order correlation coefficient, $r_s$ is shown for each correlation, along with the probability of false correlation.}
\end{figure*}

\begin{figure}[ht]
\centering
\includegraphics[width=0.7\textwidth]{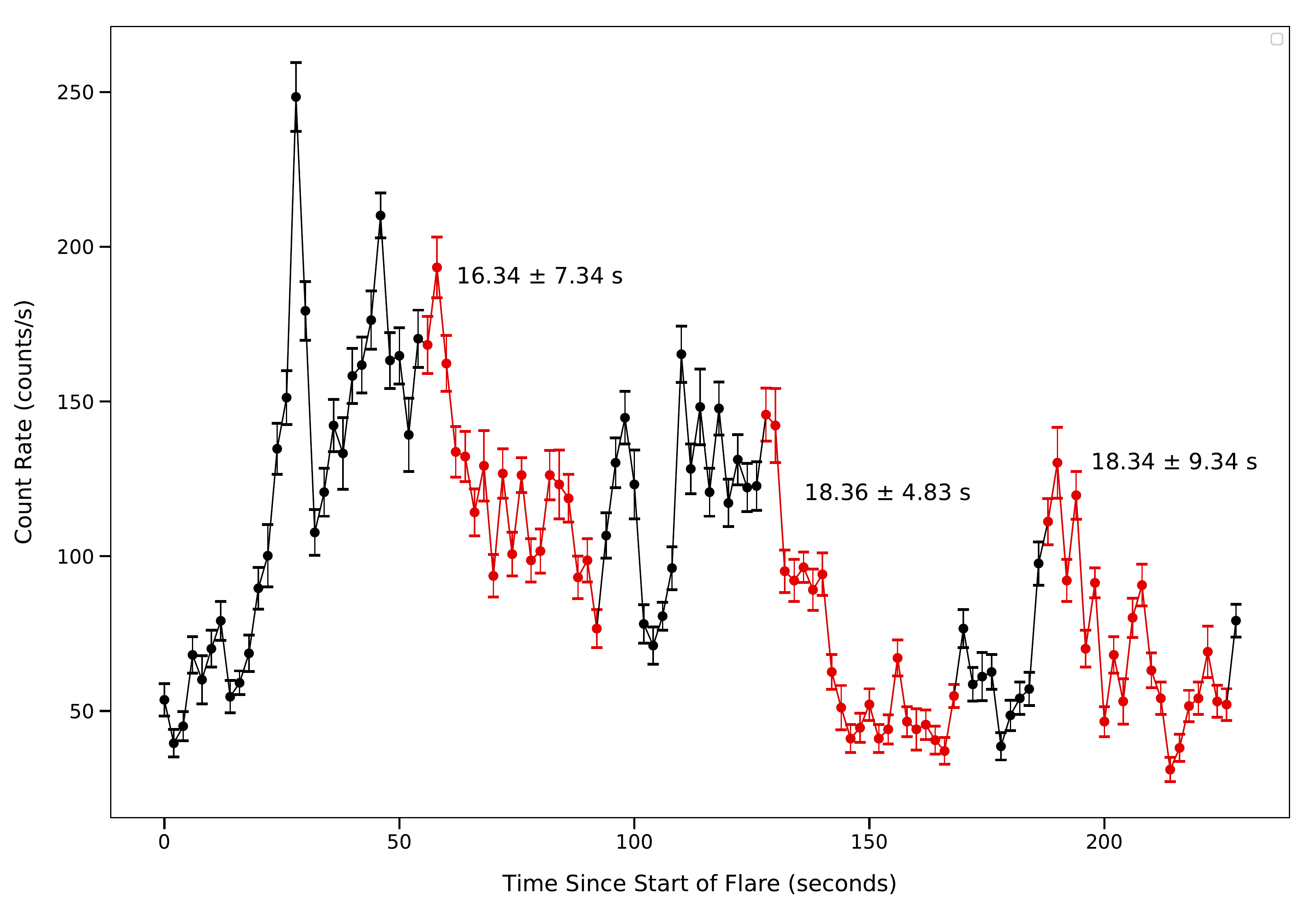}
\includegraphics[width=0.7\textwidth]{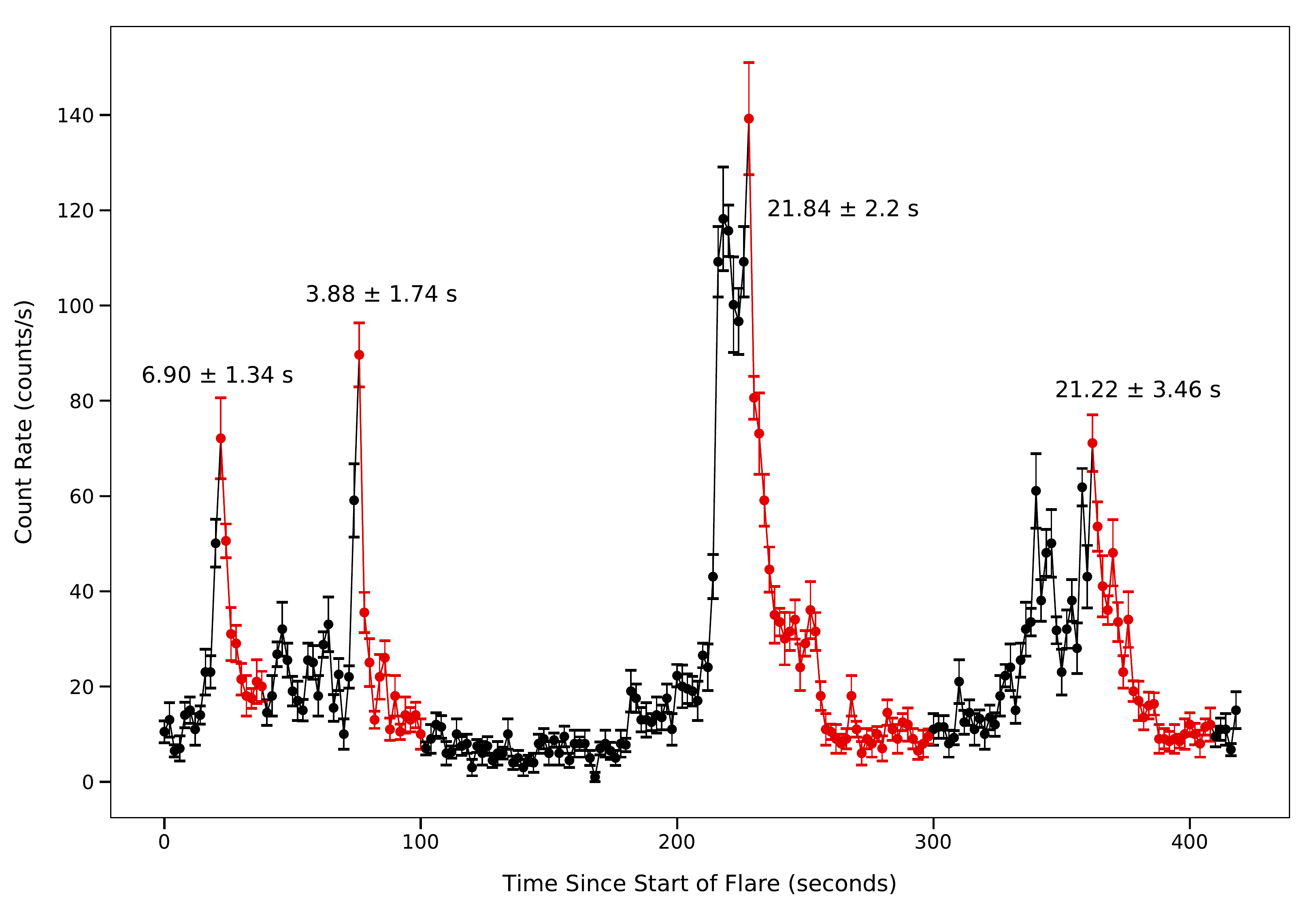}
\vspace{1mm}
\caption{Light curves of the flares captured in OBSIDs 00012178006 and 00012178037. The light curves were extracted using \texttt{XSELECT} and \texttt{XRONOS}. The light curves are plotted in units of count rate, with time bins of $\Delta t = 2$~s. The high signal-to-noise ratio allowed us to analyse multiple components of the flares separately. Data shown in red indicates the portions that were fit with an exponential function, and the text nearby reports the decay time that resulted, in seconds. Note that the x-axis indicates the time since the start of the flare, not the start of the observation that these flares were captured in.} 
\end{figure}

\begin{figure*}[ht]
\centering
\includegraphics[width = 0.47\textwidth, height=0.7\textheight]{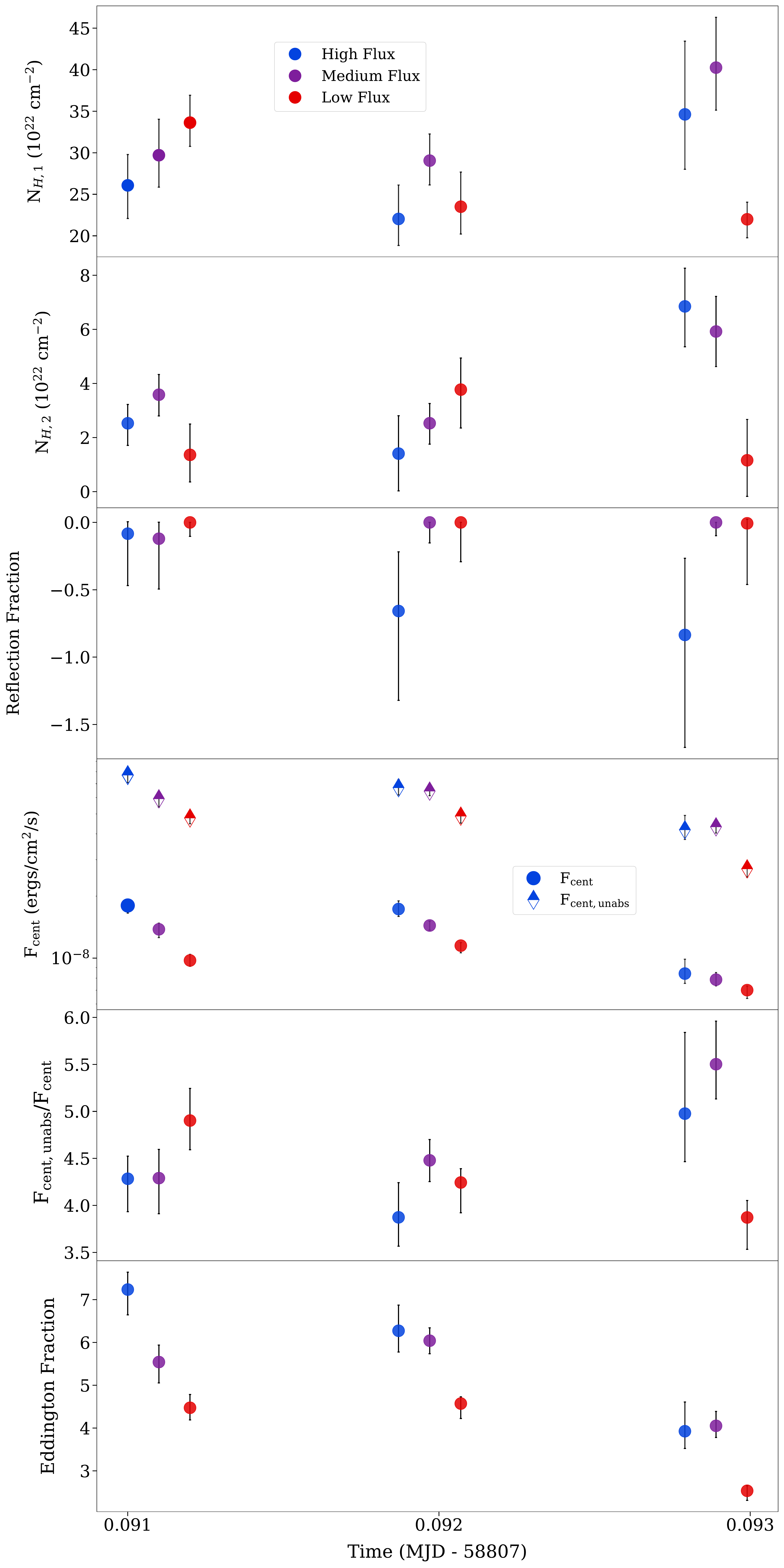}
\includegraphics[width = 0.47\textwidth, height=0.7\textheight]{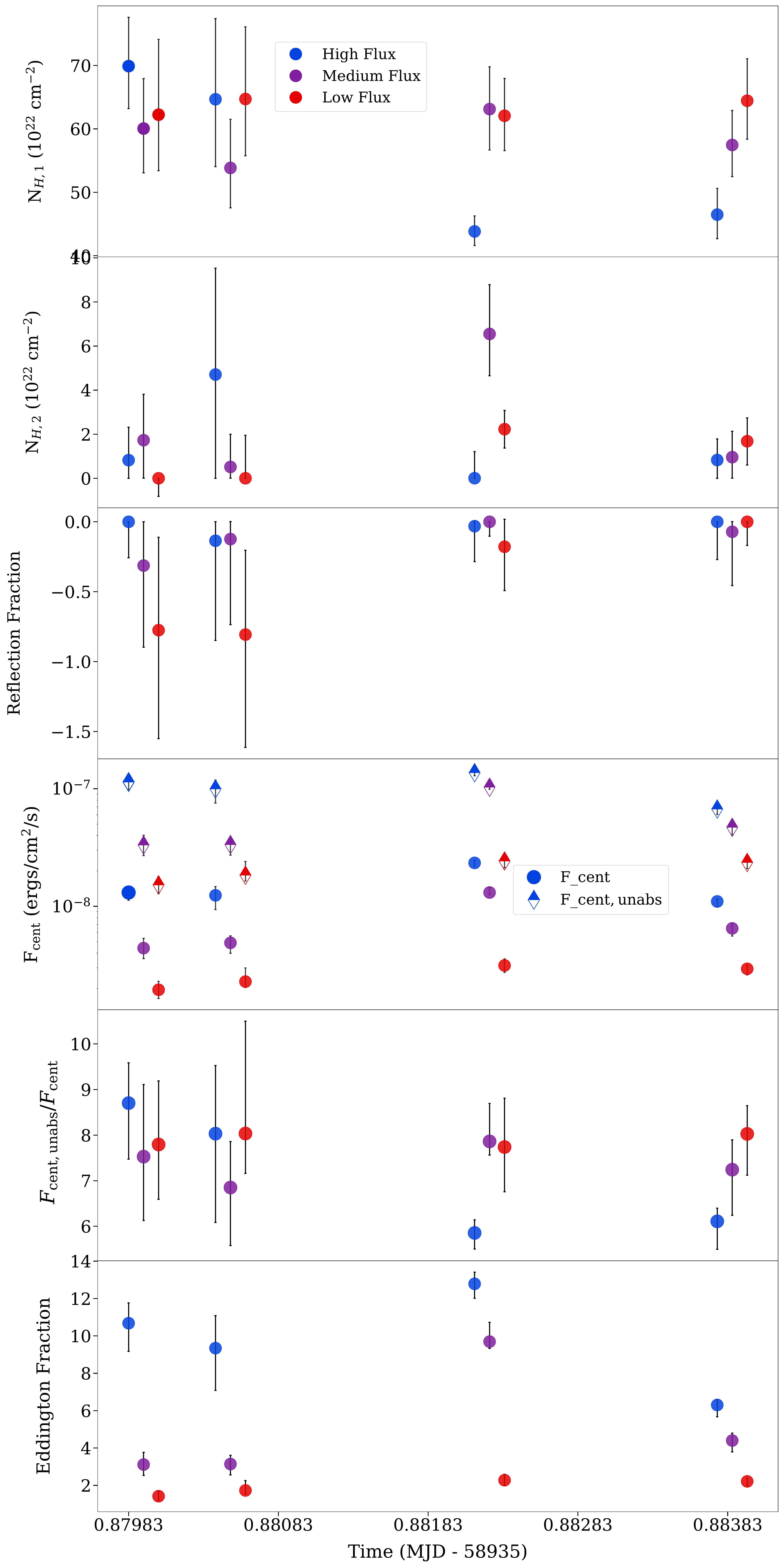}
\vspace {1mm}
\caption{Spectral parameters for flares seen in OBSID 00012178006 (left) and OBSID 00012178037 (right). The orange, purple, and red points refer to high flux, medium flux, and low flux sections of the flare, respectively. In the flux panel, the circular markers and diamond markers refer to the absorbed and unabsorbed flux of the central engine, respectively. There is no clear trend to be observed. Compared to other spectra, flares seem to have on average, lower obscuration, and a more disk-like reflector.} 
\end{figure*}

\begin{figure*}[ht]
\centering
\includegraphics[height=0.9\textheight]{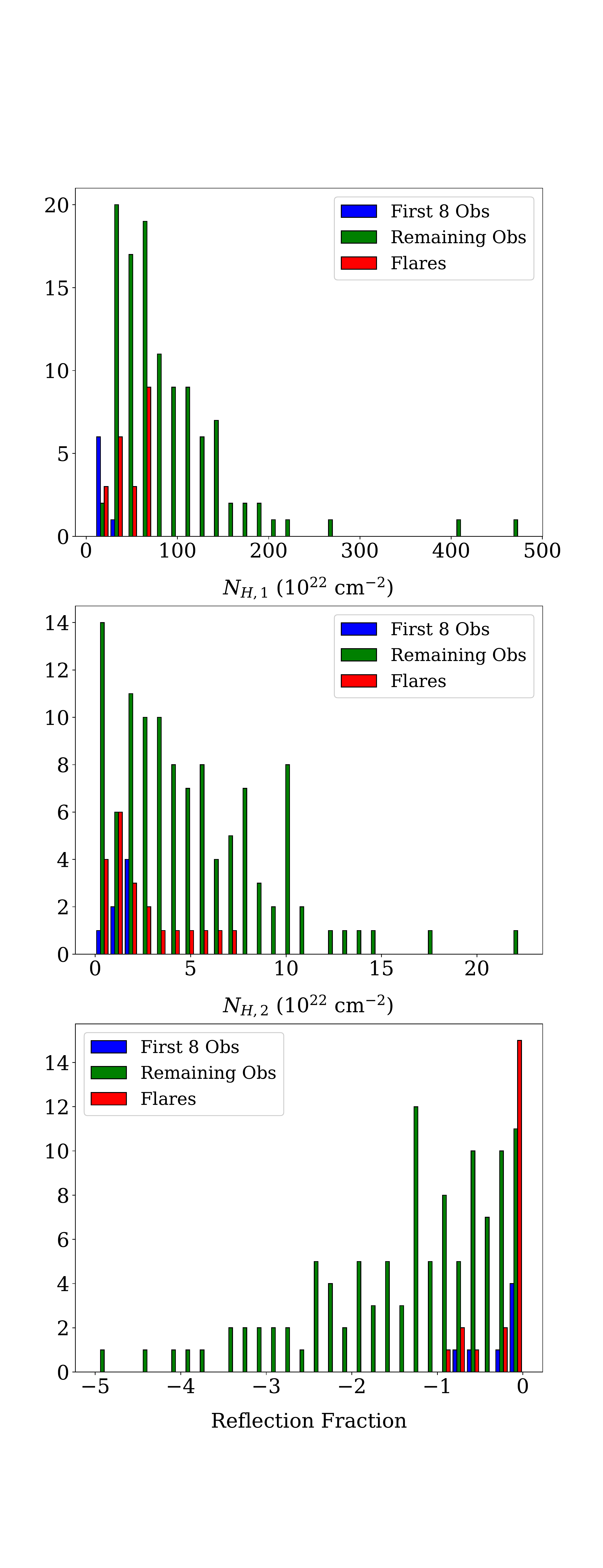}
\vspace {1mm}
\caption{Histogram of column densities and reflection fraction. The y-axis refers to the number of observations. The first few observations (which roughly correspond to the first month of the monitoring campaign) have lower obscuration values and smaller reflection fractions, similar to the flares. } 
\end{figure*}

\begin{figure*}[ht]
\centering
\includegraphics[width=0.6\textwidth]{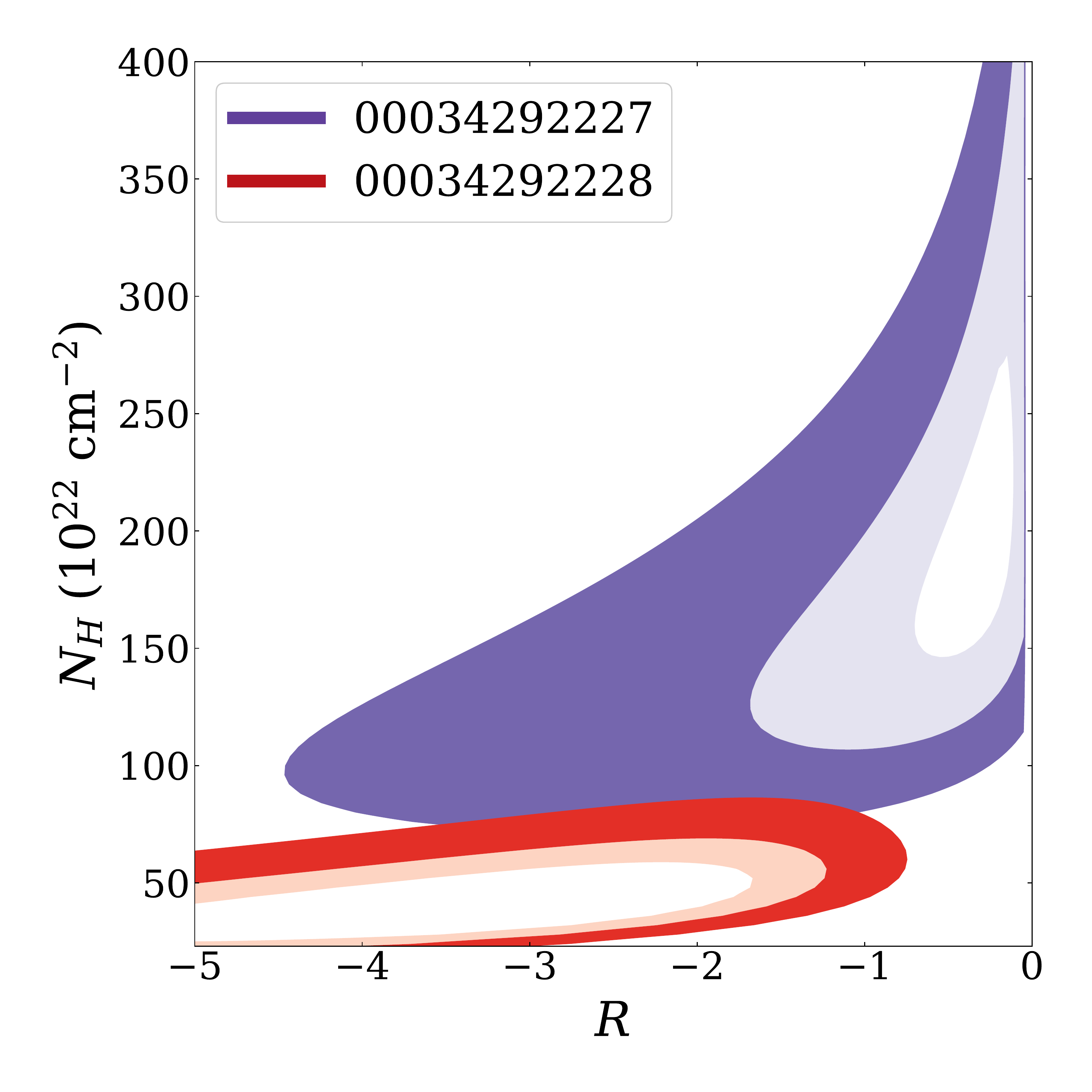}
\vspace {1mm}
\caption{Contour plot showing column density vs. reflection fraction for observations 00034292227 and 00034292228, taken one day apart. The column densities are distinctly different up to 2 $\sigma$ error bars, justifying the high values for internal obscuration returned by some of the fits.} 
\end{figure*}

\clearpage
\startlongtable
\begin{deluxetable*}{cCCC}
\tablecaption{List of \it{Swift} \rm observations}
\tablehead{
\colhead{OBSID} & \colhead{MJD - 58000} & 
\colhead{Exposure Time} & \colhead{Photon Counts}
} 
\startdata
00094149013 & 58574 & 1700 & 27000 \\
00088850001 & 58596 & 2100 & 33000 \\
00094149014 & 58597 & 1900 & 23000 \\
00034292127 & 58611 & 2000 & 7900 \\
00034292129 & 58612 & 1900 & 7900 \\
00034292131 & 58613 & 1000 & 3900 \\
00034292133 & 58614 & 870 & 2400 \\
00034292136 & 58620 & 860 & 720 \\
\enddata
\tablecomments{The table above lists the observation ID's, start time of observations (in units of MJD), exposure time, and photon counts for each observation used in our analysis. The observation campaign we used started in April 2019 and lasted for one year.}
\end{deluxetable*}

\startlongtable
\begin{deluxetable*}{cccccccc}
\tablecaption{Continuum parameters}
\tablehead{
\colhead{OBSID} & \colhead{MJD} &
\colhead{$N_{H,1}$} & \colhead{$N_{H,2}$} & \colhead{factor}  & \colhead{R}  & \colhead{C/$nu$} \\
\colhead{} & \colhead{(58000)} & \colhead{(10$^{22}$ cm$^{-2}$)} & \colhead{(10$^{22}$ cm$^{-2}$)} & \colhead{ } & \colhead{ } & \colhead{ }
} 
\startdata
00094149013 & 58574 & $10 \ \substack{+1\\ -1} $ & $0.6 \ \substack{+0.2\\ -0.2} $ & $0.6 \ \substack{+0.1\\ -0.1} $ & $0.0 \ \substack{+0\\\ -0.1 } $ & 623/626 = 0.996 \\
00088850001 & 58596 & $13 \ \substack{+1\\ -1} $ & $1.1 \ \substack{+0.2\\ -0.2} $ & $0.6 \ \substack{+0.1\\ -0.1} $ & $0.0 \ \substack{+0\\ -0.1} $ & 731/659 = 1.11 \\
00094149014 & 58597 & $17 \ \substack{+1\\ -1} $ & $1.5 \ \substack{+0.2\\ -0.2} $ & $0.5 \ \substack{+0.1\\ -0.1} $ & $0.0 \ \substack{+0 \\ -0.1} $ & 627/612 = 1.025 \\
00034292127 & 58611 & $24 \ \substack{+2\\ -2} $ & $1.7 \ \substack{+0.2\\ -0.2} $ & $0.47 \ \substack{+0.02\\ -0.02} $ & $-0.7 \ \substack{+0.2\\ -0.2} $ & 456/417 = 1.094 \\
00034292129 & 58612 & $26 \ \substack{+2\\ -2} $ & $1.7 \ \substack{+0.2\\ -0.2} $ & $0.31 \ \substack{+0.01\\ -0.01} $ & $-0.2 \ \substack{+0.2\\ -0.2} $ & 450/436 = 1.033 \\
00034292131 & 58613 & $25 \ \substack{+3\\ -2} $ & $0.9 \ \substack{+0.2\\ -0.2} $ & $0.42 \ \substack{+0.02\\ -0.02} $ & $0.0 \ \substack{+0.0\\ -0.3} $ & 271/269 = 1.009 \\
00034292133 & 58614 & $20 \ \substack{+4\\ -3} $ & $1.6 \ \substack{+0.3\\ -0.3} $ & $0.6 \ \substack{+0.1\\ -0.1} $ & $-0.8 \ \substack{+0.5\\ -0.5} $ & 144/170 = 0.851 \\
00034292136 & 58620 & $150 \ \substack{+10\\ -10} $ & $2 \ \substack{+1\\ -1} $ & $0.06 \ \substack{+0.01\\ -0.01} $ & $-0.4 \ \substack{+0.1\\ -0.1} $ & 37/38 = 0.982 \\
\enddata
\tablecomments{Continuum parameters. The table above lists the observation IDs, start time of observations (in units of MJD), column densities (in units of $10^{22} \rm cm^{-2}$), values of factor from the \texttt{const} component, and reflection fraction from the \texttt{pexmon} component. The photon reflection fraction was constrained between and $-5 \leq R \leq 0$. Errors reported are 1 $\sigma$ errors.}
\end{deluxetable*}

\startlongtable
\begin{deluxetable*}{cccccccc}
\tablecaption{Flux related parameters}
\tablehead{
\colhead{MJD}  & \colhead{F$_{\rm central}$} & \colhead{F$_{\rm scattering}$} & \colhead{F$_{\rm reflection}$} & \colhead{F$_{\rm unabsorbed}$} & \colhead{Ratio} & \colhead{ $f_{\rm Edd}$ } & \colhead{L} \\
\colhead{(58000)} & \colhead{(10$^{-10}$ $\frac{\rm erg}{\rm cm^2 s}$)} & \colhead{(10$^{-11}$ $\frac{\rm erg}{\rm cm^2 s}$)} & \colhead{(10$^{-11}$ $\frac{\rm erg}{\rm cm^2 s}$)} & \colhead{(10$^{-9}$ $\frac{\rm erg}{\rm cm^2 s}$)} & \colhead{ } & \colhead{ } & \colhead{(10$^{37}$ erg/s)}
} 
\startdata
58574 & $11\ \substack{+1\\ -1} $ & $78 \ \substack{+4\\ -4} $ & $1.4 \ \substack{+0.1\\ -0.1} \times 10^{-4} $ & $3.1 \ \substack{+0.1\\ -0.2} $ & $2.8 \ \substack{+0.1\\ -0.1} $ & $0.22 \ \substack{+0.01\\ -0.01} $ & $2.7 \ \substack{+0.1\\ -0.1} $ \\
58596 & $11 \ \substack{+1\\ -1} $ & $85 \ \substack{+3\\ -3} $ & $1.6 \ \substack{+0.1\\ -0.1} \times 10^{-4} $ & $2.9 \ \substack{+0.1\\ -0.1} $ & $2.6 \ \substack{+0.1\\ -0.1} $ & $0.20 \ \substack{+0.01\\ -0.01} $ & $2.6 \ \substack{+0.1\\ -0.1} $ \\
58597 & $8.5 \ \substack{+0.3\\ -0.3} $ & $68 \ \substack{+2\\ -2} $ & $1.4 \ \substack{+0.1\\ -0.1} \times 10^{-4} $ & $3.4 \ \substack{+0.1\\ -0.1} $ & $4.1 \ \substack{+0.1\\ -0.1} $ & $0.24 \ \substack{+0.01\\ -0.01} $ & $3.0 \ \substack{+0.1\\ -0.1} $ \\
58611 & $2.8 \ \substack{+0.1\\ -0.1} $ & $25 \ \substack{+1\\ -1} $ & $3.4 \ \substack{+0.1\\ -0.1} $ & $1.1 \ \substack{+0.1\\ -0.1} $ & $4.0\ \substack{+0.1\\ -0.1} $ & $0.079 \ \substack{+0.002\\ -0.002} $ & $1.03 \ \substack{+0.01\\ -0.01} $ \\
58612 & $3.7 \ \substack{+0.1\\ -0.1} $ & $24 \ \substack{+1\\ -1} $ & $1.5 \ \substack{+0.0\\ -0.0} $ & $1.6 \ \substack{+0.01\\ -0.01} $ & $4.3 \ \substack{+0.1\\ -0.1} $ & $0.11 \ \substack{+0.01\\ -0.01} $ & $1.42 \ \substack{+0.01\\ -0.01} $ \\
58613 & $2.6 \ \substack{+0.1\\ -0.1} $ & $23 \ \substack{+1\\ -1} $ & $1.2 \ \substack{+0.1\\ -0.1} \times 10^{-3} $ & $1.1 \ \substack{+1 \times 10^{-3}\\ -1 \times 10^{-3}} $ & $4.1 \ \substack{+0.2\\ -0.2} $ & $0.075 \ \substack{+0.003\\ -0.003} $ & $0.90 \ \substack{+0.01\\0.01} $ \\
58614 & $1.7 \ \substack{+0.1\\ -0.1} $ & $20 \ \substack{+1\\ -1} $ & $2.4 \ \substack{+0.1\\ -0.1} $ & $0.62 \ \substack{+0.01\\ -0.01} $ & $3.7 \ \substack{+0.2\\ -0.2} $ & $0.044 \ \substack{+0.002\\ -0.002} $ & $0.63 \ \substack{+0.01\\ -0.05} $ \\
\enddata
\tablecomments{Flux-related parameters. The table above lists the start time of observations (in units of MJD), the flux of the different components, the unabsorbed flux of the central engine, $F_{\rm unabsorbed} / F_{\rm absorbed}$, Eddington fraction, and source luminosity. The central engine flux, scattered flux, and reflected flux are in units of $10^{-10} \ \rm erg \ cm^{-2} \ s^{-1}$, $10^{-11} \ \rm erg \ cm^{-2} \ s^{-1}$, and $10^{-11} \ \rm erg \ cm^{-2} \ s^{-1}$ respectively. The luminosity is reported in units of $10^{37} \ \rm erg \ s^{-1}$. Errors reported are 1 $\sigma$ errors.}
\end{deluxetable*}

\begin{longrotatetable}
\begin{deluxetable*}{llllllllll}
\tablewidth{700pt}
\tabletypesize{\scriptsize}
\tablecaption{Flare Spectral Parameters}
\tablehead{ \colhead{ } & \colhead{Flux} & \colhead{MJD} & \colhead{$N_{H,1}$} & \colhead{$N_{H,2}$} & \colhead{R} & \colhead{C/$nu$} & \colhead{F$_{abs}$} & \colhead{F$_{unabs}$} & \colhead{f$_{Edd}$} \\
\colhead{} & \colhead{Level} & \colhead{} &
\colhead{(10$^{22}$ cm$^{-2}$)} & \colhead{(10$^{22}$ cm$^{-2}$)} & 
\colhead{} & \colhead{} & \colhead{(10$^{-10}$ $\frac{\rm erg}{\rm cm^2 s}$)} & \colhead{(10$^{-9}$ $\frac{\rm erg}{\rm cm^2 s}$)} & \colhead{}}  
\startdata
\multirow{3}{*}{flare 1} & high & 807.091 & $26 \ \substack{+3\\ -4} $ & $3 \ \substack{+1\\ -1} $ & $-0.1 \ \substack{+0.1\\ -0.4} $ & 322/331 = 0.974 & $180 \ \substack{+10\\ -20} $ & $77 \ \substack{+4\\ -6} $ & $7 \ \substack{+1\\ -1} $ \\
 & med & 807.091 & $30 \ \substack{+4\\ -4} $ & $4 \ \substack{+1\\ -1} $ & $-0.1 \ \substack{+0.1\\ -0.4} $ & 264/284 = 0.932 & $140 \ \substack{+10\\ -10} $ & $59 \ \substack{+4\\ -5} $ & $6 \ \substack{+1\\ -1} $ \\
 & low & 807.091 & $34 \ \substack{+3\\ -3} $ & $1 \ \substack{+1\\ -1} $ & $0.0 \ \substack{+0.0\\ -0.1} $ & 145/117 = 1.247 & $98 \ \substack{+7\\ -6} $ & $48 \ \substack{+3\\ -3} $ & $5 \ \substack{+1\\ -1} $ \\
\multirow{3}{*}{flare 2} & high & 807.092 & $22 \ \substack{+4\\ -3} $ & $1\ \substack{+1\\ -1} $ & $-0.7 \ \substack{+0.4\\ -0.7} $ & 270/217 = 1.249 & $170 \ \substack{+20\\ -10} $ & $67 \ \substack{+6\\ -5} $ & $6 \ \substack{+1\\ -1} $ \\
 & med & 807.092 & $29 \ \substack{+3\\ -3} $ & $3 \ \substack{+1\\ -1} $ & $0.0 \ \substack{ +0\\ -0.2} $ & 247/282 = 0.877 & $140 \ \substack{+10\\ -10} $ & $65 \ \substack{+3\\ -3} $ & $6 \ \substack{+0.3\\ -0.3} $ \\
 & low & 807.092 & $24 \ \substack{+4\\ -3} $ & $4 \ \substack{+1\\ -1} $ & $0.0 \ \substack{+0.0\\ -0.3} $ & 182/174 = 1.048 & $120 \ \substack{+4\\ -10} $ & $49 \ \substack{+2\\ -4} $ & $4.6 \ \substack{+0.2\\ -0.3} $ \\
\multirow{3}{*}{flare3} & high & 807.093 & $35 \ \substack{+9\\ -7} $ & $6.8 \ \substack{+1.4\\ -1.5} $ & $-0.8 \ \substack{+0.6\\ -0.8} $ & 153/152 = 1.009 & $84 \ \substack{+15\\ -9} $ & $42 \ \substack{+7\\ -4.} $ & $3.9 \ \substack{+0.7\\ -0.4} $ \\
 & med & 807.093 & $40 \ \substack{+6\\ -5} $ & $5.9 \ \substack{+1.3\\ -1.3} $ & $0.0 \ \substack{+0.0\\ -0.1} $ & 190/149 = 1.28 & $79 \ \substack{+7\\ -5} $ & $43 \ \substack{+4\\ -3} $ & $4.1 \ \substack{+0.3\\ -0.3} $ \\
\enddata
\tablecomments{The table above lists spectral parameters from the flares. The first three flares correspond to OBSID 00012178006 and the last four flares correspond to OBSID 00012178037. Shown are the time (MJD - 58000), column densities, reflection fraction, C-stat/$\nu$ values, absorbed and unabsorbed fluxes of the central engine, and Eddington fraction. The flares were fit with the same model used on the other spectra.  }
\end{deluxetable*}
\end{longrotatetable}

\newpage

\bibliography{ref}{}
\bibliographystyle{aasjournal}




\end{document}